\documentclass[a4paper,11pt]{article}
\usepackage{jcappub} 
\usepackage{lineno}

\usepackage{hyperref}
\usepackage{bm}

\arxivnumber{2508.05729} 
\title{Stable Islands of Weak Gravity}

\newcommand{\dd}{\mathrm{d}} 

\newcommand{\Mt}{$M^2$ }

\newcommand{\lcdm}{$\Lambda$CDM}
\newcommand{\mochi}{\texttt{mochi\_class}}

\author[a,b]{Linus Thummel,}
\author[a,b]{Benjamin Bose,}
\author[a,b]{and Alkistis Pourtsidou}
\affiliation[a]{Institute for Astronomy, University of Edinburgh, \\ Royal Observatory, Blackford Hill, Edinburgh, EH9 3HJ, U.K.}
\affiliation[b]{Higgs Centre for Theoretical Physics, School of Physics and Astronomy, \\ Edinburgh EH9 3FD, UK}

\emailAdd{linus.thummel@ed.ac.uk}

\abstract{
We present an exploration of the phenomenology of Horndeski gravity, focusing on regimes that produce weak gravity compared to General Relativity. 
This letter introduces a novel method to generate models of modified gravity theories that produce a specific observational behaviour while fulfilling stability criteria, without imposing a fixed parametrisation. We start from the inherently stable basis of linear Horndeski theory, implemented in the recently released Einstein-Boltzmann solver \mochi{}. The time evolution of the basis functions is designed with Gaussian processes that directly include the stability and phenomenology criteria during the generation. Here, we focus on models with weak gravity that suppress the growth of Large-Scale Structure at late times. To achieve this behaviour, we mainly focus on the design of a dynamical effective Planck mass for theories with a vanishing fifth force. We find a broad range of weak-gravity islands in Horndeski theory space. We also include additional features, like a vanishing modification to gravity at $z=0$, and extend the exploration to islands of gravity with a non-zero fifth force.
Finally, we show that replacing the \lcdm{} expansion model by the DESI $w_0w_a$CDM best fit also produces stable islands of weak gravity.
}

\begin{document}
\maketitle
\flushbottom


\section{Introduction}


This decade marks a transformative era in observational cosmology, with major galaxy surveys like DESI~\citep{DESI:2019jxc}, Euclid~\citep{Euclid:2024yrr}, Rubin's LSST~\citep{LSSTDarkEnergyScience:2018jkl}, and the Roman Space Telescope~\citep{spergel2015widefieldinfrarredsurveytelescopeastrophysics} mapping the large-scale structure (LSS) of the Universe in unprecedented detail. These missions will enable stringent tests of $\Lambda$CDM and its assumed theory of gravity, General Relativity (GR), across cosmological scales.

Motivating this effort is a set of tensions within $\Lambda$CDM, particularly in the current expansion rate $H_0$ and the clustering amplitude $S_8$, with the latter being less significant~\citep{CosmoVerse:2025txj,Wright:2025xka}. In addition, DESI has reported evidence for evolving dark energy~\citep{Collaboration-2025-DESIDR2Resultsb}, further encouraging exploration beyond standard cosmology. Among the viable theoretical alternatives are Horndeski gravity models, which introduce a scalar degree of freedom while preserving second-order field equations~\citep{Horndeski:1974wa}. Although minimal in their construction, they span a vast theory space, much of which is constrained by gravitational wave speed measurements~\citep[see][for example]{Lombriser:2015sxa,Monitor:2017mdv}, Solar System tests~\citep{Will:2014kxa}, and LSS~\citep{Gubitosi:2012hu,Bloomfield:2012ff,Frusciante-2020-EffectiveFieldTheorya}. Efficiently exploring this space of Modified Gravity (MG) requires theoretical viability and numerical stability, which has recently become more tractable through tools like \href{https://github.com/mcataneo/mochi_class_public}{\textbf{\textbf{\mochi{}}}}~\citep{Cataneo-2024-Mochi_classModellingOptimisationc}, capable of scanning the Effective field theory of Dark Energy (EFTofDE) basis while ensuring stability against ghost and gradient instabilities~\citep{Kennedy-2018-ReconstructingHorndeskiTheories,Lombriser:2018olq}. {\tt mochi\_class} also checks for exponential mode growth, further helping to restrict the viable model space. 

Rather than a blind scan, phenomenological guidance helps target viable regions. For instance, enforcing standard early-universe behaviour ensures consistency with Cosmic Microwave Background (CMB) constraints, while late-time suppression of structure growth, as suggested by the $S_8$ tension, motivates models with weaker gravity~\citep{Linder-2018-NoSlipGravity}. Similarly, increasing the inferred value of $H_0$, while recovering GR locally, would yield  physically consistent and observationally preferred theories. 

In this letter, we combine {\tt mochi\_class} with Gaussian Processes (GPs) to identify such models. GPs offer a non-parametric method to infer stable, continuous functional forms directly from data, characterizing distributions over functions through a mean and covariance function. This is optimal for navigating the EFTofDE parameter space without strong prior or parametrisation assumptions. 
\\ 
\\
\noindent {\bf Conventions:} We work in units where the Planck mass $M_\mathrm{PL}^2=1$. We normalise the scale factor $a = 1$ today. Primes denote $\ln a$ derivatives. Unless otherwise stated a \lcdm{} background with the Planck best-fit values \citep{Planck:2018vyg} is assumed. 


\section{Theory and methods} 

\subsection{Effective Field Theory of Dark Energy}  
\label{sec:EFTofDE}

Horndeski gravity describes the most general 4D scalar-tensor theory with equations of motion up to second order \citep{Horndeski:1974wa}. The EFTofDE provides a way to describe the dynamics of linear perturbations of Horndeski theory on a Friedmann-Lemaître-Robertson-Walker (FLRW) background \citep{Kennedy-2017-ReconstructingHorndeskiModels}.
It describes the background with one free function (\textit{e.g.} the Hubble expansion rate $H(a)$) and the linear perturbations with four free functions of time. The latter reduces to 3 independent functions if the propagation speed of gravitational waves $c_T$ is equal to the speed of light. This will be assumed in this paper based on gravitational wave constraints~\citep[see \textit{e.g.}][]{Lombriser:2015sxa,Monitor:2017mdv,Baker:2017hug}.

A popular basis for these functions is the \textbf{$\bm{\alpha}$-basis}~\citep{Bellini-2014-MaximalFreedomMinimum}, which provides a clear physical interpretation:
\begin{enumerate}
    \item\textbf{The value of the effective Planck Mass today $M_0$, and its running} \newline $\bm{\alpha_M} \equiv {\dd\ln M^2(a)}/{\dd \ln a}$, which tracks its rate of change.
    \item \textbf{The braiding $\bm{\alpha_B}$}, which describes the mixing of the kinetic terms of the metric and the scalar degrees of freedom.
    \item \textbf{The kineticity $\bm{\alpha_K}$}, which is linked to the kinetic energy term of the scalar field in the Lagrangian and affects LSS observables mainly on very large scales.
\end{enumerate}
A major issue of this parametrisation is that the background can become unstable to perturbations. We therefore need to ensure that our theories fulfil the following criteria \citep{Bellini-2014-MaximalFreedomMinimum, Cataneo-2024-Mochi_classModellingOptimisationc}:
\begin{align}
     M^2 &> 0\, ; \tag{tensor-mode stability} \\
     D_\mathrm{kin} &\equiv \alpha_K + \frac{3}{2} \alpha_B^2 > 0\, ; \tag{no-ghost condition} \\ 
    c_s^2 =& \dfrac{1}{D_\mathrm{kin}} \bigg[(2-\alpha_B) \left(- \dfrac{H'}{H} + \dfrac{1}{2}\alpha_B + \alpha_M \right) \nonumber  \tag{gradient stability}
    \\
    &+ \dfrac{2}{M^2}\dfrac{H'}{H} + \dfrac{3}{H^2M^2}(\rho_\phi + w_\phi \rho_\phi) + \alpha_B' \bigg]> 0 \label{eq:cs2} \, ,
\end{align}
where we have defined the de-mixed kinetic term, $\bm{D_\mathrm{kin}}$, of scalar field perturbations and their effective sound speed, $\bm{c_s}$, for wave modes ${k \rightarrow \infty}$. The terms $\rho_\phi$ and $w_\phi$ are the energy density and equation of state of the scalar field, respectively. The gradient and ghost instabilities are physically motivated and mostly affect the stability of linear perturbations in the high-$k$ regime. We do not explicitly consider the tachyonic instability of the mass terms~\citep{Frusciante-2019-RoleTachyonicInstability} but \mochi{} nevertheless ensures stability in the low-$k$ regime by testing for \textbf{mathematical instabilities}. This stability test has been adopted from the EFTCAMB implementation \citep{Hu-2017-EFTCAMBEFTCosmoMCNumerical, Cataneo-2024-Mochi_classModellingOptimisationc}. It flags theories that are not strictly theoretically forbidden but that have exponentially fast growing modes in the perturbations and are therefore ruled out by observations. The largest accepted growth rate can be tuned within \mochi{}, here we are using the default value of $\xi = 1$ (see Eq.~24 of \cite{Cataneo-2024-Mochi_classModellingOptimisationc}).

Further, we allow for superluminal sound speeds with $c_s^2>1$. \cite{Salvatelli-2016-ConstraintsModifiedGravity} have found that restricting Horndeksi theories to $c_s^2\leq 1$ significantly shrinks the stable parameter space. We were not able to identify any stable theories of weak gravity when we limited the search to models with $c_s^2=1$. While superluminal sound speeds do not necessarily lead to causality violations, achieving a consistent UV completion of these theories requires alternatives to standard methods \citep[see \textit{e.g.}][for a discussion]{Bellini-2014-MaximalFreedomMinimum,Kreisch-2018-CosmologicalConstraintsHorndeski}. This can be achieved by methods like classicalisation \citep{dvali2011uv}.

For a choice of $\alpha$-basis functional forms of time, one needs to ensure the stability criteria are met at all times. To avoid these time consuming checks, one can simply set appropriate priors in an {\bf inherently stable basis}: [$H(a)$, $\Delta M^2(a)$, $D_\mathrm{kin}(a)$, $c_s^2(a)$, $\alpha_{B_0}$]~\citep{Kennedy-2018-ReconstructingHorndeskiTheories} with $\Delta M^2(a)\equiv M^2(a)-1$.  However, we now have to solve a differential equation to obtain $\alpha_B$ (see \autoref{eq:cs2})
\begin{align}
    \alpha_B' + \left( 1 - \alpha_M + \dfrac{H'}{H}\right) \alpha_B - \dfrac{\alpha_B^2}{2}  =  - 2 \alpha_M + c_s^2 D_\mathrm{kin} 
    + \dfrac{\Delta M^2}{M^2} \dfrac{2 H'}{H} - \dfrac{3(\rho_\phi + w_\phi \rho_\phi)}{M^2H^2} \, , 
\label{eq:alphaB_ODE}
\end{align}
which is done internally by \mochi{}. It uses the quasi-static approximation (QSA), where it selectively neglects the time derivatives of scalar perturbations on small scales \citep{Cataneo-2024-Mochi_classModellingOptimisationc}.

\subsection{Parametrisations of the Basis Functions}

In order to derive data constraints on the basis functions, it has been common to assume a specific functional form for their time evolution. Popular choices for this are a power law
\begin{equation}
    \Delta M^2(a) =  (M_0^2-1) a^s \, , 
\label{eq:powerlaw}
\end{equation}
with $s$ a free parameter, or  the `$\Omega_\mathrm{DE}$' parametrisation
\begin{equation}
    \Delta M^2(a) =  (M_0^2-1) \dfrac{\Omega_\mathrm{DE}(a)}{\Omega_\mathrm{DE}(a=1)} \, .
\label{eq:Omega_DE}
\end{equation}
These examples are given for the Planck mass but they can be applied to all basis functions~\citep[see][for a review]{Frusciante-2020-EffectiveFieldTheorya}.

However, the specific constraints for each basis function depend on their selected parametrisation. This means that observational constraints are highly sensitive towards their functional form and cannot be compared between different analyses if the choice for their time evolution differs \citep{EuclidCollaboration-2025-EuclidPreparationConstraining}. 
Additionally, these parametrisations strongly restrict the parameter space and sometimes reduce it to areas of no stability~\citep{Denissenya-2018-GravitysIslandsParametrizing}, as we will show in \autoref{sec:early-time-behaviour}. 
 The basis functions $\alpha_B$ and $\alpha_M$ are the most relevant for Stage-IV LSS surveys, but their current constraints are limited to specific parametrisations~\citep{Denissenya-2018-GravitysIslandsParametrizing,Ishak-2024-ModifiedGravityConstraintsa,EuclidCollaboration-2025-EuclidPreparationConstraining}. The kineticity $\alpha_K$ cannot be constrained with current LSS observations, hence we can fix it to a constant without biasing constraints on other parameters~\citep{Bellini-2016-ConstraintsDeviationsLCDM}.

One important constraint on any MG model comes from tests of gravity on small Solar System scales that are well in agreement with GR~\citep{Will:2014kxa}. Therefore, we need to hide away any large scale modifications in these regimes. One can embed a screening mechanism into the theory which ensures GR behaviour locally~\citep[see][for a review]{Koyama:2018som}. This is typically achieved with higher-order or nonlinear terms, which are unconstrained by the basis functions of the linear theory. However, one should also ensure local constraints \textit{via} the shape of the $M^2$ curve. If we impose $M^2=1$ today then the strength of gravity returns to its GR value today, even on cosmological scales. Additionally, local measurements of the rate of change for $G$ \citep{Muller-2007-VariationsGravitationalConstant,Konopliv-2011-MarsHighResolution,Belgacem-2019-TestingNonlocalGravity} indicate that $\alpha_M \lesssim 10^{-2} $ today. Similarly, our MG models should converge to GR at early times to comply with CMB constraints~\citep{Planck:2018vyg}. Therefore we need to guarantee $\alpha_i \rightarrow 0$ (for $i=B,M,K$) and $M^2 \rightarrow 1$ for $a \rightarrow 0$.

\subsection{Weak-Gravity Theories}

MG theories can change the strength of coupling to gravity, which can be different for light and matter. Starting from the perturbed FLRW metric in the Newtonian gauge 
\begin{equation}
    \dd s^2 =   - (1+2\Psi)\dd t^2 + a^2 (1-2\Phi) \dd \bm{x}^2 \, , 
\label{eq:metric}
\end{equation}
with the gravitational potentials $\Phi, \Psi$ \citep{Bardeen:1980kt}, we obtain the modified Poisson equations for matter and light under the QSA:
\begin{align}
    k^2 \Psi &=  - 4 \pi a^2 G \mu \rho_{\rm m}\delta_{\rm m} \, ,  \\
    k^2 (\Psi + \Phi) &= - 8 \pi a^2G \Sigma \rho_{\rm m}\delta_{\rm m} \, , 
\label{eq:Poisson}
\end{align}
where $G$ is Newton's constant. The phenomenological space and time dependent functions $\mu$ and $\Sigma$ can be computed analytically from the linear basis functions under the QSA, with $\mu =\Sigma=1$ in GR. Constraints on $\mu$ and $\Sigma$ can be used to rule out subclasses of Horndeski theory~\citep{Pogosian-2016-WhatCanCosmology}, but it is harder to identify a specific model~\citep{EuclidCollaboration-2025-EuclidPreparationConstraining}. 

We will focus on the coupling strength for matter, $\mu$, that is relevant for the growth of structure. We can identify ${G_\mathrm{\rm eff} \equiv G\mu }$ as an effective gravitational constant that describes how the strength of gravity is altered by the MG model.
It is possible to express $\mu$ in terms of the linear basis functions of Horndeski theory.
The most relevant form for LSS observations is the small-scale limit in the QSA $\mu_\infty$ given by \citep{Pogosian-2016-WhatCanCosmology, Cataneo-2024-Mochi_classModellingOptimisationc}
\begin{equation}
    \mu_\infty =  \dfrac{1}{M^2} \bigg( 1 + \underbrace{\dfrac{(\alpha_B + 2 \alpha_M)^2}{2 c_s^2 D_\mathrm{kin}}}_\text{fifth force} \bigg)\, .
\label{eq:mu}
\end{equation}
Having $\mu_\infty >1$ results in stronger gravity and an enhancement in growth, while $\mu_\infty <1$ leads to weaker gravity and a suppression in growth. This can be observed as a reduction in power in the linear matter power spectrum compared to the GR spectrum. We can see from \autoref{eq:mu} and \autoref{eq:cs2} that the second term -- the fifth force -- will always be positive and enhance the strength of gravity. In order to build Horndeski models with weak gravity compared to GR, we need to have a large effective Planck mass, $M^2>1$, and an adequately small fifth force so that the total modification results in $\mu_\infty < 1$. However, fulfilling the phenomenological criteria for weak gravity \textit{and} the stability criteria of Horndeski theory is challenging. For example, \cite{Kennedy-2018-ReconstructingHorndeskiTheories} found such a model that avoids ghost and gradient instabilities but has exponentially growing modes. 

In particular, it is not possible to just alter $M^2$ while setting $\alpha_B=\alpha_K=0$, as the resulting vanishing kinetic term inadvertently leads to gradient instabilities of the theory. Furthermore, $\alpha_B$ affects relevant scales for LSS observations and disregarding it would unnecessarily restrict the parameter space for relevant phenomenologies. Similarly, we found that fixing $c_s^2=1$ shrinks the parameter space to a regime dominated by mathematical instabilities. Additionally, we found that many theories with large $M^2$ and freely varied other basis functions suffer from mathematical instabilities. By allowing $\alpha_B \neq 0$ the fifth force in \autoref{eq:mu} comes into play and potentially overpowers the suppression from a large Planck mass. This can be avoided by using the `no-slip' condition~\citep{Linder-2018-NoSlipGravity,Brush-2019-NoSlipCMB}
\begin{equation}
    \alpha_B = -2\alpha_M  \, ,
\label{eq:no-slip condition}
\end{equation}
which ensures a vanishing of the fifth force at all times. \cite{Linder-2018-NoSlipGravity} designed a \Mt shape that transitions from a GR limit at early times to a de-Sitter limit at late times given by
\begin{equation}
    \Delta M^2_\mathrm{Linder}(a) = \mu_L \dfrac{1 + \tanh [(\tau_L/2) \ln (a/a_L)]}{2}  \, , 
\label{eq:Linder-hill}
\end{equation}
with amplitude $\mu_L$,  steepness $\tau_L$, and position $a_L$ of the transition. This combined with a positive constant $\alpha_K$ and on a \lcdm{} background can produce stable weak gravity for various values of $\mu_L$, $\tau_L$ and $a_L$ \citep{Linder-2018-NoSlipGravity, Brush-2019-NoSlipCMB}.

\subsection{Gaussian Processes} 
\label{sec:GP}

We aim to generate basis curves without fixing a specific time parametrization, while incorporating phenomenological and stability constraints. To do this, we use Gaussian Process Regression (GPR) \citep{Wang-2023-IntuitiveTutorialGaussian, deJesusVelazquez-2024-NonparametricReconstructionCosmological}. Although GPR is formally parameter-free, it requires selecting a kernel and priors for its hyperparameters.

We focus on generating $\Delta M^2 \equiv M^2-1$ curves for the no-slip scenario. We later use a similar method to modify the sound speed for beyond no-slip models. Due to constraints on both the function and its derivatives, we employ the \href{https://github.com/markchil/gptools}{\textbf{\texttt{GPtools}}} package \citep{Chilenski-2015-ImprovedProfileFitting}, which allows arbitrary derivative constraints. We use the Squared Exponential (SE) kernel $k_\mathrm{SE}(x_1,x_2) = \sigma^2 \exp{-\frac{|x_1-x_2|^2}{2 l^2}}$ where $\sigma$ and $l$ control the variance and length scale of the curves.

Observational and stability constraints on $M^2$ and its derivatives are imposed in the range $-5 \leq \ln a \leq 0$. For early-time GR consistency, we set $M^2(\ln a = -5) = 1$ and $M'(\ln a = -5) = 0$ -- with small errors to allow for gradual convergence --  and $(M^2)'' \approx 0$ for gradient stability. To model weak gravity, we push towards $M^2 > 1$ in the range $-3 \leq \ln a \leq -1$, permitting some tolerance as long as the value remains above unity. If agnostic about nonlinear screening mechanisms, we can conservatively opt to impose $M^2 \approx 1$ and $\alpha_M \approx 0$ today in order to respect Solar System tests. The no-ghost condition is automatically satisfied by choosing a positive and constant  $\alpha_K$.

 The ultimate goal of this workflow is to identify stable models that produce a specific phenomenology and then map them to a fully covariant theory which is far more instructive for model building and connecting to more fundamental descriptions of Nature. In this work, we're showing how to use Gaussian processes to comprehensively identify islands in theory space that fulfil the phenomenological criterium and pass all stability tests. In an upcoming work, we will show how we can obtain analytical expressions for these theories using symbolic regression \citep{Cranmer-2023-InterpretableMachineLearning}. This opens up the opportunity to constrain the stable islands with observational data. Using the mapping provided in \cite{Lombriser:2018olq}, we can derive the subclass of covariant descriptions corresponding to the symbolically regressed theory, and thereby close the link from observational trend to fundamental description of cosmology and gravity.


\section{Model Selection and Results}

\subsection{Effective-Planck-Mass Function Generation}
\label{sec:early-time-behaviour}

\begin{figure}
        \centering
        \includegraphics[width=0.68\columnwidth]{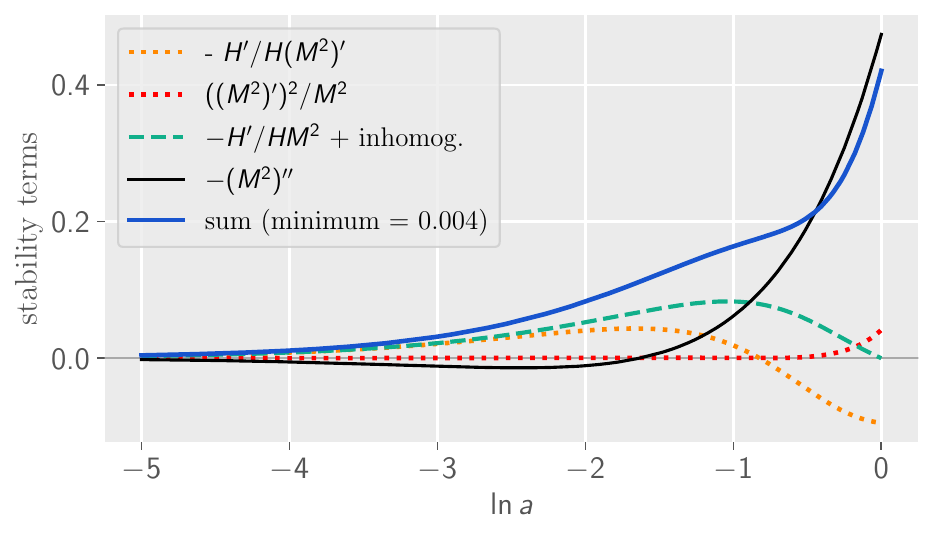} 
  \caption{\textbf{Testing gradient stability:} Terms entering   \autoref{eq:cs2-terms-noSlip} for the local-value weak-gravity model. To achieve a stable theory with $c_s^2>0$ we require the sum of all the plotted terms (in blue) to be positive at all times.}
  \label{fig:cs2-terms}
\end{figure}

In this section, we present the generation of effective Planck mass curves that yield stable weak gravity. We start with models obeying the no-slip condition while fixing the background and the value of $\alpha_K$. This approach is based on the $\alpha$-basis, but we will later switch to a direct use of the stable basis for beyond-no-slip models.

\begin{figure*}
        \centering
        \includegraphics[width=\linewidth]{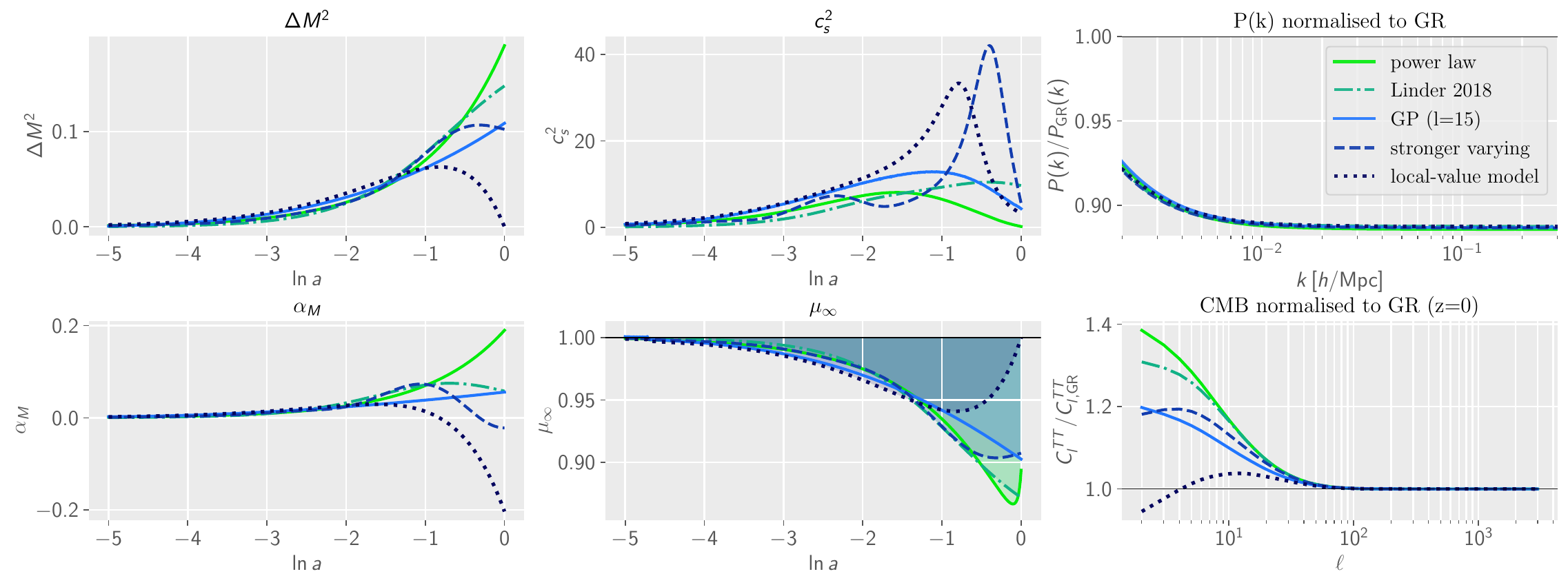} 
  \caption{\textbf{$\bm{M^2}$ curves of stable no-slip models that produce weak gravity:} Selection of $\Delta M^2 \equiv M^2-1$ basis functions that lead to stable weak gravity with the no-slip condition $\alpha_B = - 2 \alpha_M$. The models are built with a \lcdm{} background and $\alpha_K=10^{-2}$. All curves are stable against ghost, gradient, and mathematical instabilities as tested in \mochi{}. The amplitude of each $\Delta M^2$ curve is scaled to produce $\sim 12 \%$ suppression of the linear matter power spectrum (top right panel) compared to GR at $z=0$. They are shown with the corresponding no-slip quantities and the generated suppression.    
  }
  \label{fig:M2_collection}
\end{figure*}

A key challenge is avoiding the gradient instability, which we address by translating the sound speed constraint into conditions on $M^2$ and its derivatives. By substituting $\alpha_B$ in \autoref{eq:cs2} with the no-slip condition, we get a constraint for $M^2$ shapes, with a clear condition for gradient stability:
\begin{align}
   0 < \dfrac{M^2}{2} c_s^2 D_\mathrm{kin}  = & - (M^2)'' + \dfrac{((M^2)')^2}{M^2} - \dfrac{H'}{H}(M^2)' - \dfrac{H'}{H}M^2 \nonumber \\ 
   & + \underbrace{\dfrac{H'}{H} + \dfrac{3}{2 H^2}(\rho_\phi + w_\phi \rho_\phi)}_\text{inhomog.}   \, .
\label{eq:cs2-terms-noSlip}
\end{align}
The various terms are plotted for a stable weak-gravity $M^2$-example in \autoref{fig:cs2-terms}. In \autoref{fig:M2_collection}, it is shown with other  $M^2$ curves that are generated using our method assuming \lcdm{} and constant $\alpha_K = 10^{-2}$. 
One challenge for gradient stability in no-slip models is the early-time behaviour, implemented as $M^2 \to 1$ and $(M^2)' \to 0$ for $\ln a \to -5$. The $- (M^2)''$ and $(M^2)'$ terms become dominant as the inhomogeneous and $M^2$ terms cancel each other at early times. This makes it essential that \texttt{GPtools} supports higher-derivative constraints.

Early-time stability is challenging, but our focus is on late-time phenomenology relevant to Stage-IV surveys. We can therefore begin with an analytical form that ensures stable evolution as $a \rightarrow 0$, and introduce GP modifications only at late times.\footnote{We include a pure GP curve (with $\sigma = 3$, $l=15$) in \autoref{fig:M2_collection} to show that early-time stability can also be achieved just from GPs.} This also guarantees smooth convergence beyond the finite interval where GP curves are defined. We test three functional forms for stability, \textit{i.e.}, whether they satisfy $c_s^2 > 0$ at all times. If so, they can serve as bases for late-time GP modifications. As shown by \citep{Linder-2018-NoSlipGravity}, the shape in \autoref{eq:Linder-hill} leads to stable no-slip models; we confirm this for various parameters and show the case with $\mu_L = 0.2$, $\tau_L = 1.5$, $a_L = 0.5$ in \autoref{fig:M2_collection}. The same holds for the power law form in \autoref{eq:powerlaw}, with stability verified,\textit{ e.g.}, for $M_0^2 = 1.19$, $s = 1$. In contrast, the $\Omega_\mathrm{DE}$ form in \autoref{eq:Omega_DE} fails to yield a stable model. This highlights the sensitivity of stability to the chosen time parametrisation, reinforcing the need to explore a broad range of functional forms rather than fixating on any single one.

Using a sign analysis of \autoref{eq:cs2-terms-noSlip}, we can assess the stability for various possible features of $M^2$ curves. Maxima 
are generally allowed if $\alpha_K$ is large enough to prevent divergence in $1/D_\mathrm{kin}$. We show a GP curve with strong variations\footnote{The stronger varying GP curve ($\sigma = 1$, $l = 0.8$) is superimposed on a power law base to ensure early-time stability. The same is the case for the GP of the local-value curve ($\sigma = 3$, $l = 5$).} as an example in \autoref{fig:M2_collection}. Minima, however, are typically unstable: with $(M^2)'=0$ and $(M^2)'' > 0$, stability would require larger $M^2$ and late-time $\rho_\phi$.

Finally, we can also impose GR recovery today in accordance with more local tests of gravity. However, \autoref{eq:cs2-terms-noSlip} shows that it is not possible to find a weak-gravity no-slip model that starts from GR at early times and also fulfils both $M^2=1$ and $\alpha_M=0$ today (`local-convergence') and is stable for a \lcdm{} background. The sound speed always drops to negative values at late times. However, we can generate a stable theory with $M^2=1$ today (`local-value'), which is shown in \autoref{fig:M2_collection}. We will present a local-convergence model that is breaking the no-slip condition in \autoref{sec:beyond-no-slip}.

\subsection{Constructing No-Slip Models}

We will now construct no-slip theories and beyond based on the designer GP \Mt curves for weak gravity. Assuming that we input a fixed background evolution (\textit{e.g.} \lcdm{} or $w_0w_a$CDM) and a fixed $\alpha_K>0$, we proceed as follows:
\begin{enumerate}
\item choose a background evolution for $H(a)$;
\item design an \Mt shape for a specific phenomenology (here for weak gravity with $\mu_\infty <1$) and the physical stabilities listed in \autoref{eq:cs2}. (See \autoref{sec:early-time-behaviour} for the detailed procedure);
\item compute $\alpha_M$;
\item get $\alpha_{B}$ and $\alpha_{B0}$ from \autoref{eq:no-slip condition};
\item combine the chosen $\alpha_K = 10^{-2}$ with $\alpha_{B}$ to get $D_\mathrm{kin}$ ($D_\mathrm{kin}>0$ is fulfilled by construction as $\alpha_K>0$);
\item compute the consistent no-slip sound speed $c_{s}$ with \autoref{eq:cs2}, checking that $c_{s}^2>0$ holds at all times;
\item for beyond no-slip: modify $c_{s}^2$ with a GP and recheck $c_s^2>0$;
\item feed all stable basis functions  [$\Delta M^2$, $D_\mathrm{kin}$, $c_{s}^2$, $H (a)$, $\alpha_{B0}$] into \mochi{} and test for mathematical instabilities. 
\end{enumerate}

\autoref{fig:M2_collection} shows the no-slip quantities for the $M^2$ shapes presented previously. All normalised quantities are shown relative to a \lcdm{} model without modifications to gravity. Each curve is scaled to yield $\sim 12\%$ suppression in the linear matter power spectrum at $z = 0$ relative to GR, motivated by an interacting dark energy model with similar growth suppression effects \citep{Pourtsidou-2016-ReconcilingCMBStructure}. As expected for no-slip models, $\mu_\infty < 1$ at all times. While the redshift evolution of $\mu_\infty$ varies between models, the $k$-dependence of the power suppression remains the same. Slightly enhanced large-scale power reflects the breakdown of the QSA. None of the models shifts the CMB TT peaks, but they differ in their large-scale Integrated-Sachs-Wolfe (ISW) effects. The ISW amplitude and suppression strength can be tuned by adjusting the $\Delta M^2$ amplitude.

All curves correspond to stable configurations. The $\Delta M^2$ shapes in \autoref{fig:M2_collection} can be rescaled and shifted while preserving stability and suppression. Since suppression strength tracks the amplitude of $\Delta M^2$, we can construct models consistent with observations, and we plan to constrain such models with CMB and LSS data in future works.

\begin{figure*}
        \centering
        \includegraphics[width=\linewidth]{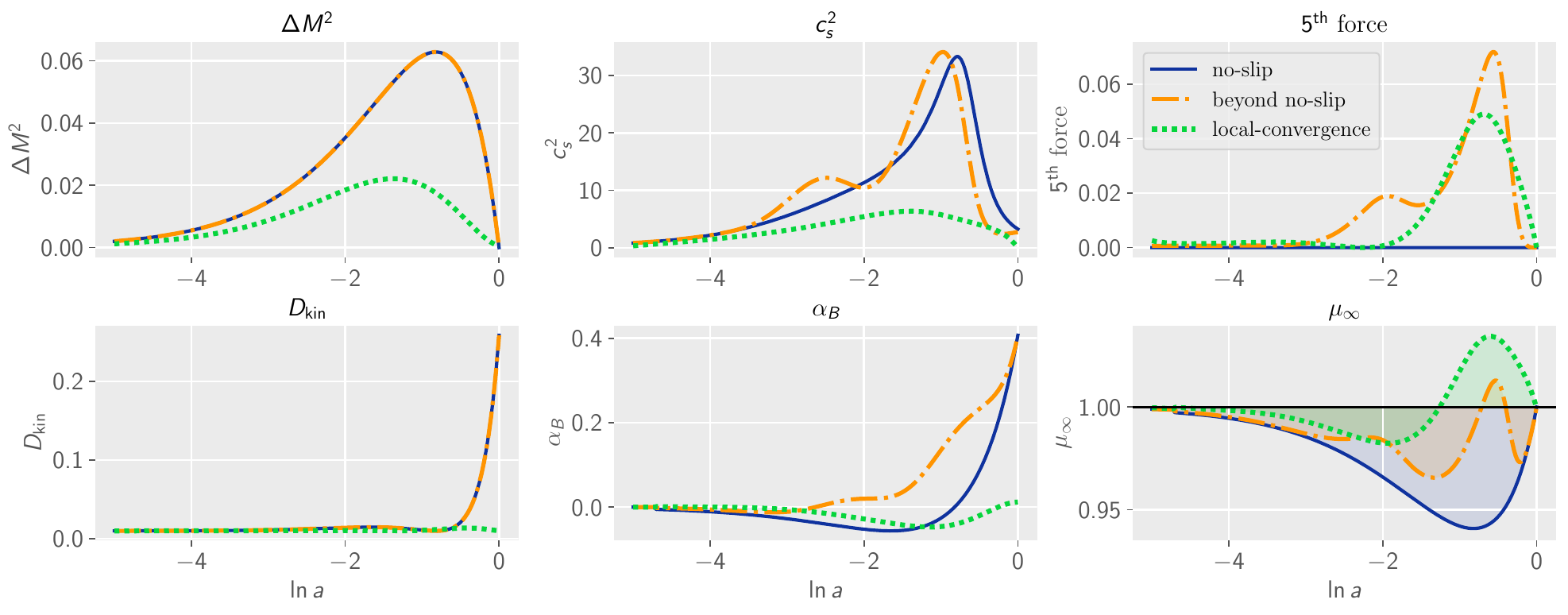} 
  \caption{\textbf{No-slip and beyond-no-slip models:} Stable basis functions and the corresponding suppression $\mu_\infty$ from \autoref{eq:mu} for the local-value \Mt model. Blue curves represent the no-slip model (with $\alpha_B = -2 \alpha_M$).  We compare it to the beyond no-slip model (orange) where the sound speed $c_s^2$ is modified with a GP curve -- inducing a fifth force -- while $M^2$, $D_\mathrm{kin}$ and $\alpha_{B0}$ are kept unchanged. The dotted lines represent a beyond-no-slip local-convergence model with $M^2\approx1$ and $\alpha_M\approx0$ today.}
  \label{fig:mu_M2_example}
\end{figure*}

\subsection{Beyond-No-Slip Models}
\label{sec:beyond-no-slip}
If we allow models to break the no-slip condition (\autoref{eq:no-slip condition}), we can freely modify the sound speed for every \Mt even for a fixed background as long as they remain stable. Starting from the no-slip models in the previous section, we will explore scenarios beyond no-slip by adding GP modifications to their sound speed.
This is illustrated in \autoref{fig:mu_M2_example}, where we consider the local-value model with $M^2(a=1) = 1$ from the previous section without loss of generality, keeping $D_\mathrm{kin}$ and $\alpha_{B0}$ fixed to their no-slip values. The modified sound speed alters $\alpha_B$ via \autoref{eq:alphaB_ODE}, introducing a non-zero fifth force (right panel of \autoref{fig:mu_M2_example}) and increasing $\mu_\infty$, potentially reducing suppression or even producing strong gravity phases.

A wide range of sound speed variations can yield stable weak gravity models when GP curves for $c_s^2$ are carefully chosen. This is aided by the designer $M^2$ curves optimized for suppression. The sound speed modifications must preserve early-time convergence, maintain $c_s^2 > 0$, and limit the fifth force to avoid prolonged enhancement. Using a stable basis allows us to impose gradient stability checks directly during GP generation. We allow for superluminal sound speeds with $c_s^2>1$, as they are beneficial for stability and weaker gravity. The product $c_s^2 D_\mathrm{kin}$ has a strong influence on the strength of gravity of the model. It acts as a denominator in the fifth force of \autoref{eq:mu}, hence a small sound speed would lead to enhanced gravity in models with non-zero slip.

\begin{figure*}
        \centering
        \includegraphics[width=0.49\linewidth]{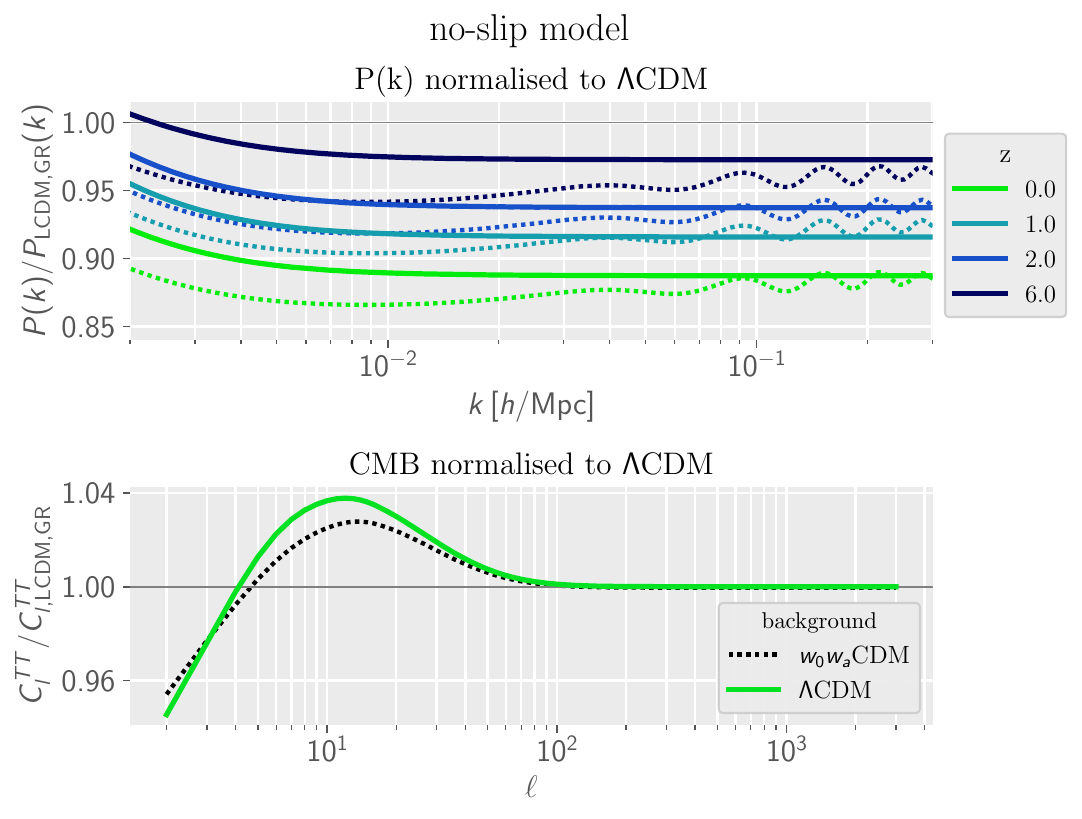} 
        \includegraphics[width=0.49\linewidth]{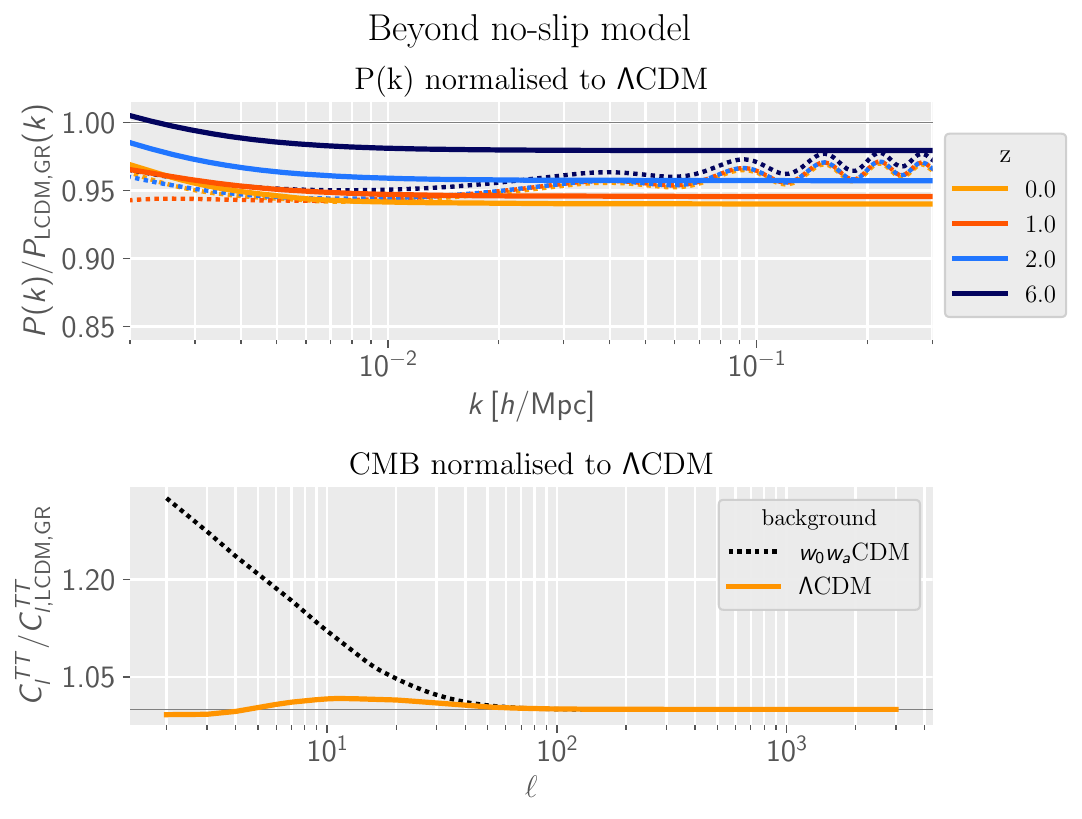} 
  \caption{\textbf{Suppression for \lcdm{} and $\bm{w_0w_a}$CDM backgrounds:} Effects of the no-slip (left side) and beyond-no-slip (right side) models for the local-value Planck mass shown in \autoref{fig:mu_M2_example} on the CMB TT power spectrum and linear matter power spectrum. We show the \mochi{} outputs of the MG models for two different backgrounds that are always normalised to the GR model with a \lcdm{} background. Dotted lines represent a MG model with $w_0w_a$CDM background for the DESI best-fit values ($w_0 = -0.752$, $w_a = -0.86$). Coloured solid lines are used for MG models with a \lcdm{} background.}
  \label{fig:Pk_CMB_M2_example}
\end{figure*}

To analyse the suppression by weak-gravity models in more detail, we show the linear matter power spectrum across redshifts in \autoref{fig:Pk_CMB_M2_example}. Beyond-no-slip models exhibit reduced suppression compared to no-slip, as the fifth force offsets the effects of large $M^2$: periods with $\mu_\infty < 1$ are balanced by phases with $\mu_\infty > 1$. We also observe a reduced ISW effect. We further show a beyond-no-slip local-convergence model with $M^2\approx1$ and $\alpha_M\approx0$ today. It is produced by shifting and rescaling the (partly negative) no-slip sound speed of a local-convergence $M^2$ curve.

\subsection{Evolving Background}
\label{sec:w0wa}

All equations in this letter apply to a general background $H(a)$, not necessarily \lcdm{}. Here we show results when replacing the \lcdm{} background in the no-slip and beyond-no-slip local-value models with the best-fit $w_0w_a$CDM background from the latest DESI DR2 release \citep{Collaboration-2025-DESIDR2Resultsb}: $w_0 = -0.752$ and $w_a = -0.86$.

\autoref{fig:Pk_CMB_M2_example} compares the linear matter and CMB TT power spectra for the same stable basis functions but different backgrounds. The normalisation spectrum is the GR model with a \lcdm{} background and the same cosmological parameters. Note that we fix the angular scale at last scattering $\theta_\star$. Hence, varying the background evolution results in different $H_0$ values that change the scale of the Baryon Acoustic Oscillations.  
Switching to a $w_0w_a$CDM background (dotted curves) still leads to stable models with a suppression of growth. However, the background has a significant influence on the amplitudes and shapes of the suppression and ISW effect. The change is even more pronounced in the beyond-no-slip scenario.   This provides a promising basis for further studies of background phenomenology within Horndeski theory.


\section{Conclusions}
We have presented a novel method for constructing stable Horndeski theories that realize specific phenomenology, focusing on models that suppress structure growth with $\mu_\infty < 1$. Our framework builds upon the inherently stable basis and its implementation in the Boltzmann solver \mochi{}, enabling a parameter-free exploration of theory space. Physical stability conditions (\autoref{eq:cs2}) and phenomenological requirements are encoded as constraints in Gaussian Process Regression, ensuring all generated models are stable and testable within \mochi{}.

To achieve suppression, we designed the effective Planck mass $M^2$, requiring $M^2 > 1$ for weak gravity. However, the fifth force (\autoref{eq:mu}) can counteract this, so we began with no-slip models, which eliminate the fifth force by construction. These models allow maximal suppression and simplify the stability analysis.
We identified various stable $M^2$ shapes, with or without analytic early-time convergence to GR, including cases which return to GR locally ($M^2 = 1$ today). Rescaling these curves yielded further stable models, forming islands of weak gravity within no-slip theory space. By modifying the sound speed with GP curves, we explored the beyond no-slip regime, reintroducing the fifth force and expanding the viable model space. With carefully constructed modifications, these beyond-no-slip models also remain stable and suppress growth. In future work, Symbolic Regression \citep{Cranmer-2023-InterpretableMachineLearning} can be used to parametrize these stable islands efficiently and identify classes of observationally viable covariant theories.

Phenomenologically, the resulting models exhibit similar suppression in the matter power spectrum and only affect the ISW effect without shifting the primary CMB peaks. Replacing the \lcdm{} background with a $w_0w_a$CDM model still yields stable, weak gravity solutions, suggesting broader potential for exploring dynamical backgrounds in Horndeski theory. In upcoming work, we will show how we can use Symbolic Regression to obtain analytical expression for the identified islands and map them to a fully covariant theory.


\acknowledgments

L.T.'s research is supported by a PhD studentship from the Science and Technology Facilities Council (STFC).
B.B. is supported by a UKRI Stephen Hawking Fellowship (EP/W005654/2). A.P. is a UK Research and Innovation Future Leaders Fellow [grant MR/X005399/1]. 
We wish to thank Matteo Cataneo for the detailed discussions on this topic. We are grateful to him and Emilio Bellini for the publicly available Boltzmann code \mochi{} for the stable basis. 
We thank Mark Chilenski for developing the GP code \texttt{GPtools} that allows for arbitrary derivative constraints.
We also thank the organisers and participants of the ``Testing gravity with the Large-Scale Structure of the Universe'' focus week in Trieste for inspiring discussions. 
For the purpose of open access, the authors have applied a Creative Commons Attribution (CC BY) licence to any Author Accepted Manuscript version arising from this submission.


\bibliographystyle{JHEP}
\bibliography{biblio.bib}

\end{document}